\documentclass[aps,prd,12pt,preprint,tightenlines,unsortedaddress,
amsfonts,amssymb,amsmath,byrevtex,showpacs,nofootinbib]{revtex4-1}
\usepackage{amssymb,amsmath}
\usepackage{latexsym}
\usepackage{graphicx}
\usepackage{geometry}                
\usepackage{color}
\DeclareMathOperator\arcosh{arcosh}

\newcommand{\R}{\mathbb R}

\newcommand{\ket}[1]{|\kern.3ex#1\kern.3ex\rangle}
\newcommand{\bra}[1]{\langle\kern.3ex #1 \kern.3ex|}
\newcommand{\scalar}[2]{\langle\kern.3ex #1 \kern.3ex|\kern.3ex#2\kern.3ex\rangle}
\newcommand{\norm}[1]{\|\kern.3ex#1\kern.3ex \|}

\def\bsb{\boldsymbol{\beta}}

\begin{document}
\title[Toda]{Semi-Classical quantisation of 3-particles Toda lattice augmented\\
Application to the Mixmaster anisotropy Hamiltonian}

\author{Herv\'{e} Bergeron}
\email{herve.bergeron@universite-paris-saclay.fr} \affiliation{Univ Paris-Saclay, ISMO, UMR 8214 CNRS,
91405 Orsay, France}
\affiliation{APC, Universit\'e Paris Diderot, Sorbonne Paris Cit\'e, 75205 Paris
Cedex 13, France}

\author{Ewa Czuchry}
\email{ewa.czuchry@ncbj.gov.pl} \affiliation{National Centre for Nuclear Research, Pasteura 7,
02-093 Warszawa, Poland}

\begin{abstract}

Usual approaches to quantisation of a 3-Toda  lead to numerical calculations requiring many steps that can be time consuming to insure their reliability. In order to reduce as much as possible the numerical part of the EKB quantisation procedure, and then to ease numerical calculations, we  propose a reformulation of the mathematical framework with more adapted variables. The resulting  equations and procedure might be easily implemented in a short Mathematica code. This more explicit framework will be useful for studying quantum Toda-Bianchi IX models in quantum cosmology where the true Bianchi IX anisotropy Hamiltonian can be approximated by  a 3-particle Toda system.
\end{abstract}

\maketitle

\tableofcontents

\section{Introduction}

In our recent papers \cite{qb9f,qmixV2} we have demonstrated that the quantum behaviour of a refined model of the earliest Universe, namely the so-called ``Mixmaster universe", might be approximated by a 3-particle Toda  system, usually arriving in solid state physics.
The classical dynamics of the ``Mixmaster universe" (or Bianchi IX model) was studied by C. Misner in the canonical formalism \cite{cwm} at the end of the sixties. It involves a Hamiltonian that is formally identical to that of a particle moving in a 3D Minkowski spacetime in a time-dependent, exponentially steep and triangle-like potential.  Many mathematical studies have been devoted to the  classical  evolution  of  the  system   and  have  led  to  some  important  results on the asymptotic behaviour, the non-integrability or the chaotic behaviour (see e.g.  \cite{Uggla1}--\cite{Conte}).

On  the  other  hand,  the  quantum  behaviour  of  Mixmaster  remains  not entirely  understood despite many interesting studies \cite{qb9f, qb9montani, qb9craig, qb9damour, qb9prd2016, qb9univ}. The  difficulty lies mainly  in the very different possible dynamical regimes induced by the elaborate spatial dependence of the anisotropy potential.
We (with our co-authors) have shown in \cite{qmixV2}  that the classical Mixmaster anisotropy potential can be viewed as a difference between two 3-particle Toda potentials, thus paving the way to a possible analytically solvable approximation for the problem of the quantum version of the Bianchi IX model (see Sec. 5 for more details). This approximation should be valid  in the unexplored  region  in-between  the known harmonic and steep-wall approximations.\footnote{The latter corresponding respectively to the large volume and low anisotropy excitation level or to the  small volume and high anisotropy excitation level.}

The Toda lattice system is a one-dimensional chain consisting  of equal masses  interacting via  exponential forces. This system was shown to be completely integrable \cite{Flaschka} --\cite{Ford} as it has  as many constants of motion as the numbers of   degrees of freedom.
The quantisation of a Toda lattice was first performed by Gutzwiller \cite{Gutzwiller} who formulated a systematic recursive way of constructing the eigenfunctions and  explicitly constructed them  for $N=2,3$ and $4$ periodic Toda systems. However it took much time until numerical results appeared \cite{Mutsuyama} and \cite{Isola}, as they needed calculational power and time. Canonical quantisation was performed after direct diagonalization of the Hamiltonian, and results were classified with respect to representation of the permutation group  $S_3$ under which the Hamiltonian is invariant. It was found that the numerically obtained  eigenvalues fulfil Gutzwiller's quantisation conditions. Furthermore the semiclassical Einstein-Keller-Brillouin method of quantisation (EKB quantisation) was proved to provide results in good agreement with exact ones \cite{Mutsuyama}, even for the first levels. However the usual mathematical formulation of the EKB quantisation in the case of 3-Toda system involved several implicit functions that seem to impose a pure numerical approach with not simple entangled steps. Therefore, at first sight, it seems difficult to export  this procedure in a straightforward way in the more complex framework of Bianchi IX.

The aim of this note is to prove that the mathematical formulation of the 3-Toda EKB quantisation can be significantly eased by a new parametrisation of the problem.

The outline of the paper is as follows. In Sec. II we recall the main features of the classical 3-body Toda lattice and its known formalism for EKB semi-classical quantisation. This section allows to introduce all the notations useful for the remainder.  In Sec. III we introduce a new parameterisation for the Toda system which we use in Sec. IV for new parameterisations of the action integrals which turn to be easily calculated. In Sec. V we discuss the possible application of the obtained procedure to quantum Mixmaster with the anisotropy Hamiltonian where a pure Toda potential is used but with restored necessary dependences. We conclude in Sec. VI.

\section{Classical Toda lattice and its  semi-classical quantization}

\subsection{Classical formulation}
  {The Toda lattice system is a one-dimensional chain consisting  of equal masses  interacting via  exponential forces.
There are two types of those, an open lattice one  and a periodic one. }The main difference is that in the periodic lattice the first and the last particles are coupled whereas in   the open lattice they are not. The Hamiltonian for the periodic Toda system is like for a system of $N$ equal-mass particles interacting via exponential potential:
\begin{equation}
H=\frac12\sum_{k=1}^N p_k^2 +\sum_{k=1}^N e^{-(q_k-q_{k+1})},
\end{equation}
with periodicity condition $q_0 \equiv q_N$ and $q_1\equiv q_{N+1}$,   {where $q_i$ are generalised positions and $p_i$ their corresponding conjugate momenta. }

The simplest nontrivial  periodic crystal  is  the periodic 3-particle Toda system with Hamiltonian as follows:
\begin{equation}\label{ham3per}
H=\frac12\left(p_1^2+p_2^2+p_3^2\right)+e^{-(q_1-q_3)}+e^{-(q_2-q_1)}+e^{-(q_3-q_2)}.
 \end{equation}
The equations of motion for this system may be written as Lax's equation \cite{Flaschka}:
\begin{equation}
\frac{dL}{dt}=\left[M, L\right],
\end{equation}
where matrices $L$ and $M$  read  as follows

\begin{equation}
L:=\begin{bmatrix}
b_1 & a_1 & a_3 \\
 a_1 & b_2 & a_2 \\
a_3 &a_2 & b_3
 \end{bmatrix}, \ \
 M:=\frac12 \begin{bmatrix}
 0 & a_1 & -a_3 \\
 -a_1 & 0 & a_2 \\
 a_3 &-a_2 & 0
 \end{bmatrix}.
\end{equation}
Elements of the symmetric matrix $L$ and the skew one $M$ are functions of positions and momenta of a Toda system:
\begin{equation}\label{abdef}
a_i:=\frac12e^{(q_i-q_{i+1})/2}\ \textrm{and}\ b_i:=\frac{p_i}2\ \textrm{where}\ i={1,2,3},\ \textrm{with}\ q_0\equiv q_3,\ q_4\equiv q_1.
\end{equation}
  {The matrices $L$ and $M$ form a so called Lax pair, therefore the eigenvalues of $L$ and also  the coefficients of its characteristic polynomial $A_i$, are constants of motion:}
\begin{equation}
\det (2\mu I-2L)\equiv (2\mu)^3+A_1(2\mu)^2+A_2(2\mu)+A_3-2
\end{equation}
In the center of mass system $P:=p_1+p_2+p_3\equiv 0$     {those coefficients simplify to}:
\begin{equation}
A_1=-P=0,\ A_2=\frac12 P^2-H=:-E,\
\end{equation}
\begin{equation}
A_3=\Pi_{i=1}^3  p_i-\sum_{i=1}^3 p_ie^{q_{i+1}-q_{i+2}}\equiv 8\Pi_{i=1}^3  b_i-8\sum_{i=1}^3 b_ia^2_{i+1}=:A.
\end{equation}
Therefore, conserved quantities $A_1$ and $A_2$ have physical interpretations, respectively the total momentum $P$ and the energy of the system $E$. The third conserved quantity $A$ does not have such explicit simple physical meaning.

Hamiltonian \eqref{ham3per} is invariant under the transformations of the  dihedral group $D_3$. The dihedral group $D_N$ is  the group of symmetry of the $N$-sided regular polygon. The particular group $D_3$  has two kinds of representations, two  one-dimensional representations $A_1$, $A_2$ and  a two-dimensional representation $E$.

  { The Toda lattice system is integrable even if the Hamiltonian is not separable, i.e. we cannot separate its variables into the explicit ones. However, due to  integrability, we can rewrite the Hamiltonian by means of the canonical transformation in terms of the action-angle variables $(I_i, \theta_i)$. For our $n=3$ Toda chain we have $i=1,2$. Moreover, it is possible to introduce the canonical conjugate variables $(\mu_i,\nu_i)$, where $i=1,2$. The action-angle variables arise in integrable systems and can be written as
\begin{equation}
I_i:=\oint \nu_i(\mu_i) d\mu_i,\label{action-integral}
 \end{equation}
where integration is performed along the closed loop trajectory in phase space, here over a period of motion. It was calculated in \cite{Toda} that for a Toda lattice conjugated momenta (such that Poisson brackets fulfil $\{\nu_i,\mu_j\}=\delta_ij$) are following:
\begin{equation}
\nu_{1,2} =  2 \ln | \frac12( \Delta(\mu_i)\pm\sqrt{\Delta(\mu_i)^2-4}|,
\end{equation}
where
\begin{equation}
\Delta(\mu):=\det(2\mu I-2L)+2.
\end{equation}
In the the center of mass system $P:=p_1+p_2+p_3\equiv 0$ this function reads as:
\begin{equation}
\Delta(\mu) =8 \mu^3 - 2 E \mu +A,
\end{equation}
The variables $\mu_i$ are defined only on a limited interval specified by the inequality $|\Delta(\mu)|\geq 2$ and so also action integrals $I_i$ are taken over that regions. The expression $|\Delta(\mu)|\geq 2$ } is clearly a 3rd order polynomial and thus has 2 intervals where inequality $|\Delta(\mu)|\geq 2$ holds.
Since $E \ge0$, $\Delta(\mu)$ admits a relative maximum for $\mu = -\mu_m$  and a relative minimum for $\mu = +\mu_m$ with $\mu_m  = \sqrt{\dfrac{E}{12}}$.
\begin{equation}
\begin{tabular}{|c|ccccccc|}
\hline
$\mu$ & $- \infty$ & $ $ & $- \mu_m$ & $ $ &$ + \mu_m$ & $ $ & $+ \infty$ \\
\hline
$ $ & $ $ & $ $ & $\Delta(-\mu_m)$ & $ $ & $ $ & $ $ & $+ \infty$\\
$\Delta(\mu)$ & $ $ & $ \nearrow$ & $ $ & $ \searrow $ & $ $ & $ \nearrow$ & $ $ \\
$ $ & $- \infty $ & $ $ & $ $ & $ $ & $ \Delta(\mu_m)$ & $ $ & $ $\\
\hline
\end{tabular}
\end{equation}
In order that the inequality $|\Delta(\mu)|\geq 2$  holds on a non-zero measure interval of $\mu$
 each of the two equations $\Delta(\mu) = 2$ and $\Delta(\mu) = -2$ must have three distinct real  roots. This is only possible if $\Delta(-\mu_m) > 2$ and $\Delta(\mu_m) < -2$. This holds for following values of $E$ and $A$:\begin{equation}
\label{conds}
E > 3 \text{ and } |A| < \dfrac{2}{3 \sqrt{3}} E^{3/2} - 2.
\end{equation}
Therefore, these two conditions have to be fulfilled to develop semi-classical quantisation.

  {  We call $\mu_1^+ < \mu_2^+ < \mu_3^+$ the solutions of $\Delta(\mu)=+2$ and $\mu_1^- < \mu_2^- < \mu_3^-$ the three solutions of $\Delta(\mu)=-2$. The canonical conjugate momenta $\nu_i^+$ and $\nu_i^-$ such that $\{\mu_i^+,\nu_j^+\}=\delta_{ij}= \{\mu_i^-,\nu_j^-\}$ are given by:
 \begin{equation}
\left.
\begin{array}{c}
\nu_i^+ =  2 \ln | \frac12( \Delta(\mu_i^+)+\sqrt{\Delta(\mu_i^+)^2-4})|, \\
\nu_i^-=  2 \ln | \frac12( \Delta(\mu_i^-)-\sqrt{\Delta(\mu_i^-)^2-4}) |.
\end{array}\right. 
\end{equation}}

\subsection{EKB quantization}
\label{secEKBproblems}
  {It was shown in \cite{Fergusson} that for a 3 particle Toda lattice
 the corresponding  actions $I_1(E,A)$ and $I_2(E,A)$ read as
\begin{align}
\label{actionsdef}
I_1(E,A)& = 4 \int_{\mu_1^+}^{\mu_2^+} \arcosh \dfrac{|\Delta(\mu)|}{2} d \mu\;, \nonumber \\
I_2(E,A) &= 4 \int_{\mu_2^-}^{\mu_3^-} \arcosh \dfrac{|\Delta(\mu)|}{2} d \mu \;,
\end{align}}
where $\arcosh(x) = \ln(x+\sqrt{x^2-1})$ for $x \ge1$.
As we see, the energy $E$ and the conserved quantity $A$, are given implicitly in terms of the  above actions $I_i$. Furthermore, from \eqref{conds} the range of possible values of $A$ is dependent of the value of $E$ and obviously  the boundaries of integrals are not defined in an explicit way as functions of $E$ and $A$.

Now the  semiclassical quantisation performed through EKB (Einstein-Keller-Brillouin) formulation can be obtained as follows:  inverting (at least formally) equations \eqref{actionsdef}, the classical Hamiltonian can be expressed in terms of actions $I_i$, and the semiclassical quantum energies are finally given by the substitutions $I_i\mapsto (n_i +1/2 ) h$ in the expression of the Hamiltonian. In other terms EKB quantisation consists in finding the solutions $E_{n_1,n_2}$ and $A_{n_1,n_2}$ of the system of equations
\begin{align}
\label{sys1}
I_1(E_{n_1,n_2}, A_{n_1,n_2}) &= 2 \pi \hbar \left(n_1+ \frac{1}{2} \right),\nonumber\\
  I_2(E_{n_1,n_2}, A_{n_1,n_2})& = 2 \pi \hbar \left(n_2+ \frac{1}{2} \right).
\end{align}
for integer values of $n_1$ and $n_2$.  \\
  {Equations \eqref{actionsdef} might be a bit simplified by  using the definitions of $\arcosh $ and then performing integration by parts:
\begin{align}
\label{actionsdef1}
I_1(E,A)& = 4 \int_{\mu_1^+}^{\mu_2^+} \log\left|\frac12 \left(\Delta(\mu)+\sqrt{\Delta(\mu)^2-4}\right) \right| d \mu= \nonumber \\
&=-4 \int_{\mu_1^+}^{\mu_2^+}\frac{\mu \Delta'(\mu) }{\sqrt{\Delta(\mu)^2-4}}d \mu\;,\nonumber\\
I_2(E,A) &= 4 \int_{\mu_2^-}^{\mu_3^-} \log\left|\frac12 \left(\Delta(\mu)+\sqrt{\Delta(\mu)^2-4}\right) \right| d \mu= \nonumber \\
&=-4  \int_{\mu_2^-}^{\mu_3^-} \frac{\mu \Delta'(\mu) }{\sqrt{\Delta(\mu)^2-4}}d \mu\;,
\end{align}}

Obviously the solutions can only be found numerically and the difficulties of the procedure lie mainly in:
\begin{itemize}
\item[(a)] the non-complete independence of parameters $E$ and $A$ due to conditions \eqref{conds},
\item[(b)] the implicit definitions of the boundaries of integrals $I_1$ and $I_2$ in terms of $E$ and $A$,
\item[(c)] the dependence in $E$ and $A$ of the length of integration domains (lengths that can be very large),
\item[(d)]  the final numerical procedure to solve the system \eqref{sys1}.
\end{itemize}
To obtain the sought quantities $E_{n_1,n_2}$ and $A_{n_1,n_2}$ the authors in \cite{Isola} change in a controlled manner the values of $E$ and $A$, calculating the corresponding actions and retaining those values of $E$ and $A$ which satisfy conditions \eqref{sys1}. The author of \cite{Mutsuyama}  used the results of a direct canonical quantisation as initial values for solving the system \eqref{sys1} via the simplex algorithm. Both approaches are numerically demanding,   {specially as we see in \eqref{actionsdef1}  the integrated function is divergent at both integral limits. That demands a very accuracy of numerical integration! }\\

Our approach presented in this paper consists in removing first the difficulties (a), (b) and (c) on analytical level by a new parametrisation of the problem, and performing numerical calculations only for the final step (d) but in much simplified settings.

\section{New parametrisation for 3-Toda system}
The main point of our approach consists in introducing a new parametrisation of the problem in terms of two completely independent parameters $(\alpha, \theta) \in ]0,+\infty[ \times ]0, \pi[$ in place of the usual dynamical constants $E$ and $A$ constrained by the conditions \eqref{conds}. These two parameters  $(\alpha, \theta)$ are defined such that:
\begin{equation}
\label{newparam}
\left.
\begin{array}{c}
E := 3 \cosh^{4/3} \alpha ,\\
A := - 2 \cos \theta \sinh^2 \alpha.
\end{array}\right.
\end{equation}
  {Due to internal properties of the hyper/trygonometric functions the} conditions \eqref{conds} are  automatically fulfilled with a one to one correspondence,   {as $\forall \alpha, \theta : \in \R \cosh \alpha \ge1$ and $|\cos\theta|\le 1$.}  This solves the first point (a) mentioned at the end of the previous section \ref{secEKBproblems}.

  {The second problem is that we have to deal with a problematic integral  with no predefined bounded length. In order to solve that} let us rescale the variable $\mu$ which appears in $\Delta(\mu)$ and defining a new variable $\nu$ as follows
\begin{equation}
\mu: = \nu \mu_m
\end{equation}
where $\mu_m$ is the value of $\mu$ where $\Delta(\mu)$ reaches its relative extremal value, namely $\mu_m = \sqrt{E/12}$. With the new parametrisation  \eqref{newparam} we have:
\begin{equation}
\label{eqn:mum}
\mu_m = \frac{1}{2} \cosh^{2/3} \alpha.
\end{equation}
Finally, the expression $\Delta(\nu \mu_m)$ reads as:
\begin{equation}
\label{newparam1}
\Delta(\nu \mu_m) = (\cosh^2 \alpha) \, (\nu^3 - 3 \nu) - 2 \cos \theta \sinh^2 \alpha.
\end{equation}
In the remainder we will prove that the maximal range of $\nu$ in integrals will be reduced to $\nu \in [-2,+2]$, solving the point (c) mentioned in section \ref{secEKBproblems}.

\subsection{New parametrisation of the roots $\mu_1^+$ and $\mu_2^+$}
In this section we focus on the point (b) of section \ref{secEKBproblems}, i.e. how to obtain an explicit mathematical expression of the boundaries of integrals \eqref{actionsdef} in terms of the new parameters $(\alpha, \theta)$. Let us recall that the $\mu_i^+$ involved in the boundaries of integrals are solutions of $\Delta(\mu_i) = 2$. Using the new parametrisation of \eqref{newparam1} we obtain in terms of $\nu$ the equation
\begin{equation}
\nu^3 - 3 \nu = 2 \left(1 - 2 \sin^2 \frac{\theta}{2} \tanh^2 \alpha \right).
\end{equation}
  {In the Appendix \ref{appendix} we have described a procedure of finding explicit solutions of the 3rd order polynomial of the form $\nu^3 - 3 \nu = 2 \cos \Phi$ leading to the explicit solutions in $\nu$. In order to apply this procedure to the above equation }  let us introduce a new parameter $\Phi^+\in ]0, \pi[$ such that
\begin{equation}
1 - 2 \sin^2 \frac{\theta}{2} \tanh^2 \alpha = \cos \Phi^+.
\end{equation}
This is equivalent to the equation
\begin{equation}
\sin^2 \frac{\theta}{2} \tanh^2 \alpha = \sin^2 \frac{\Phi^+}{2}.
\end{equation}
Because $\alpha >0$, $\theta/2, \Phi^+ /2 \in ]0, \pi/2[$ implying $\sin (\theta/2),\ \sin (\Phi^+/2) > 0$, we can simplify this equation as follows
\begin{equation}
\sin \frac{\theta}{2} \tanh \alpha = \sin \frac{\Phi^+}{2},
\end{equation}
and then
\begin{equation}
\label{defPhi}
\Phi^+ = 2 \arcsin \left[ \sin \frac{\theta}{2} \, \tanh \alpha \right].
\end{equation}
Introducing the parameters $\nu_i^+ $ such that $\mu_i^+  = \nu_i^+ \mu_m$ and using the solutions \eqref{solspoly} of appendix \ref{appendix}, we find that the sought values $\nu_1^+$ and $\nu_2^+$ needed for the definition of $I_1$ (which is now a function of $\alpha$ and $\theta$ in place of $E$ and $A$) are explicitly:
\begin{equation}
\label{eqn:nu}
\nu_1^+ = -2 \cos \frac{\pi- \Phi^+}{3} \quad < \quad \nu_2^+ = -2 \cos \frac{\pi+\Phi^+}{3},
\end{equation}
where $\Phi^+$ defined in \eqref{defPhi} is an explicit function of $\alpha$ and $\theta$.

\subsection{New parametrisation of the roots $\mu^-_2$ and $\mu^-_3$}
Similarly let us recall that the $\mu^-_i$ are solutions of $\Delta(\mu_i) = -2$. Using again the  parametrisation of \eqref{newparam1} we obtain in terms of $\nu$ the equation
\begin{equation}
\nu^3 - 3 \nu = - 2 \left(1 - 2 \cos^2 \frac{\theta}{2} \tanh^2 \alpha \right).
\end{equation}
Now let us introduce a new parameter $\Phi^-$ similar to the one used previously, namely we would like to find $\Phi^- \in ]0, \pi[$ such that
\begin{equation}
- \left(1 - 2 \cos^2 \frac{\theta}{2} \tanh^2 \alpha \right)= \cos \Phi^-.
\end{equation}
This is equivalent to
\begin{equation}
\cos^2 \frac{\theta}{2} \tanh^2 \alpha = \cos^2 \frac{\Phi^-}{2}.
\end{equation}
Because $\alpha >0$, $\theta/2, \Phi^- /2 \in ]0, \pi/2[$ and then $\cos (\theta/2),\ \cos (\Phi^-/2) > 0$, we can simplify this equation as
\begin{equation}
\cos \frac{\theta}{2} \tanh \alpha = \cos \frac{\Phi^-}{2},
\end{equation}
and then we have
\begin{equation}
\label{defPhiprime}
\Phi^- = 2 \arccos \left[ \cos \frac{\theta}{2} \, \tanh \alpha \right].
\end{equation}
Introducing the parameters $\nu^-_i$ such that $\mu^-_i  = \nu^-_i \mu_m$ and using the solutions \eqref{solspoly} of appendix \ref{appendix}, we find that the sought values $\nu^-_2$ and $\nu^-_3$ needed for the definition of $I_2$ (which is now a function of $\alpha$ and $\theta$ in place of $E$ and $A$) are explicitly:
\begin{equation}
\label{eqn:nuprime}
\nu^-_2 = -2 \cos \frac{\pi+\Phi^-}{3} \quad < \quad \nu^-_3 = 2 \cos \frac{\Phi^-}{3},
\end{equation}
where $\Phi^-$ defined in \eqref{defPhiprime} is an explicit function of $\alpha$ and $\theta$.

\section{New parametrisation of the actions $I_1$ and $I_2$}
Using the definitions \eqref{actionsdef} of $I_1$ and $I_2$ and using the change of variable $\mu = \nu \mu_m$ in the integrals, we end with the formula
\begin{equation}
\label{actionsdef1}
\left.
\begin{array}{c}
I_1(\alpha,\theta) = 4 \mu_m \int_{\nu_1^+}^{\nu_2^+} \arcosh \dfrac{|\Delta(\mu_m \nu)|}{2} d \nu \\
I_2(\alpha,\theta) = 4 \mu_m \int_{\nu^-_2}^{\nu^-_3} \arcosh \dfrac{|\Delta(\mu_m \nu)|}{2} d \nu
\end{array}\right. \;,
\end{equation}
where $\mu_m$ given in \eqref{eqn:mum}; $\nu_1^+$, $\nu_2^+$ given in \eqref{defPhi}, \eqref{eqn:nu}; $\nu^-_2$, $\nu^-_3$ given in \eqref{defPhiprime} \eqref{eqn:nuprime} and the function $\nu \mapsto \Delta(\nu \mu_m)$ given in \eqref{newparam1} are explicit functions of $\alpha$ and $\theta$.   {We may observe that in our new parameterization the integrated functions in the expressions for $I_1$ and $I_2$ do not exhibit any singular behavior!}

For the final step (i.e. solving the EKB system \eqref{sys1} for a given pair $(n_1, n_2)$), these integrals \eqref{actionsdef1} must be  computed numerically. But since the function $\nu \mapsto \arcosh \Delta(\nu \mu_m)$ of Eq.\eqref{newparam1} has no singular behaviour on each interval and the intervals $[\nu_1^+, \nu_2^+]$, $[\nu^-_2, \nu^-_3]$ have a maximal length of 4, the numerical estimates are easy and reliable.

The search of semi-classical quantised energies can now be done numerically without much effort. We used a program Mathematica running on a laptop. After quick numerical calculations using build-in functions
we obtained solutions  in terms of $\alpha$ and $\theta$. Finally,  we needed only to use the definition of $E$ and $A$ in terms of $\alpha$ and $\theta$  in \eqref{newparam} to go back to the sought quantities.

As a proof of efficiency of our procedure, we can compare the values obtained with our method (with few seconds of computation on a laptop) with the ones calculated  in \cite{Mutsuyama} (where $\hbar=1$). The results are summarised in the table \eqref{results}: the agreement is perfect.

\begin{equation}
\label{results}
\begin{tabular}{|c||c|c||cc||cc|}
\hline
Symmetry & $ $ & $ $ & Matsuyama \cite{Mutsuyama} & $ $ & Our code & $ $\\
\hline
A or E & $n_1$ & $n_2$ & $E$ & $A$ & $E$ & $A$\\
\hline
A & 0 & 0 & 4.7748 & 0 & 4.7748 & 0\\
A & 1 & 1 & 8.5854 & 0 & 8.5854 & 0\\
A & 3 & 0 & 10.8558 & 9.2294 & 10.8558 & 9.2293\\
A & 4 & 4 & 21.9378 & 0 & 21.9378 & 0\\
A & 7 & 1 & 22.6452 & 29.0562 & 22.6452 & 29.0562\\
\hline
E & 1 & 0 & 6.6686 & 2.4110 & 6.6686 & 2.4110\\
E & 0 & 2 & 8.7002 & -5.4897 & 8.7002 & -5.4897\\
E & 1 & 3 & 12.8280 & -6.9356 & 12.8279 & -6.9356\\
E & 5 & 1 & 17.5336 & 16.6194 & 17.5336 & 16.6194\\
\hline
\end{tabular}
\end{equation}
Remark on a particular situation: for $n_1=n_2$ (which belongs to the symmetry ``A'') it is possible to prove directly that  $A=0$, which means in terms of our variables that $\theta = \pi/2$. Therefore, only the energy $E$ (or the parameter $\alpha$) is unknown. Then it is better from a numerical point of view to solve a unique equation either the one involving $I_1$ or the one in $I_2$, imposing by hand $\theta=\pi/2$.

\section{Application to the semi-classical quantisation of Toda-Bianchi IX anisotropy Hamiltonian}
\subsection{Bianchi IX general framework}
Let us first recall  Hamiltonian formulation of the Bianchi type IX model.  The respective Hamiltonian constraint in the Misner variables reads \cite{cwm}:
\begin{equation}\label{con}
H_{B9} =\frac{{N}e^{-3\Omega}}{24}\left(\frac{2\kappa}{\mathcal{V}_0}\right)^2\left(-p_{\Omega}^2+\mathbf{ p}^2+36\left(\frac{\mathcal{V}_0}{2\kappa}\right)^3\frak{n}^2e^{4\Omega}[V(\bsb)-1]\right)\,,~~(\Omega,p_{\Omega},\bsb, \mathbf{p})\in\mathbb{R}^6,
\end{equation}
where $\bsb:=(\beta_{+},\beta_-)$, $\mathbf{p}:= (p_+,p_-)$, $\mathcal{V}_0=\frac{16\pi^2}{\frak{n}^3}$ is the fiducial volume, $\kappa=8\pi G$ is the gravitational constant, ${ N}$ is the non-vanishing and otherwise arbitrary lapse function. The variable $\Omega$ describes the isotropic geometry, whereas $\beta_{\pm}$ describe distortions to the isotropic geometry and are referred to as the anisotropic variables.  \\
In what follows we set  $\frak{n}=1$ and $2\kappa=\mathcal{V}_0$. The spacetime variables used in eq. (\ref{con}) have the following metric interpretation:
\begin{equation}\Omega=\frac13\ln a_1a_2a_3,~~\beta_+=\frac16\ln\frac{a_1a_2}{a_3^2},~~\beta_-=\frac{1}{2\sqrt{3}}\ln\frac{a_1}{a_2}~.\end{equation}

The Hamiltonian constraint (\ref{con}) is a sum of the isotropic and anisotropic parts, $\mathrm{C}=-\mathrm{C}_{iso}+\mathrm{C}_{ani}$, where (up to a factor)
\begin{align}\label{condec1}
\mathrm{C}_{iso}&=p_{\Omega}^2+36e^{4\Omega},\\
\label{condec2}\mathrm{C}_{ani}&=\mathbf{p}^2+36e^{4\Omega}V(\bsb)\, ,
\end{align}
and the anisotropy potential $V(\bsb)$ reads as:
\begin{equation}\label{b9pot}
V(\bsb) = \frac{e^{4\beta_+}}{3} \left[\left(2\cosh(2\sqrt{3}\beta_-)-e^{- 6\beta_+}
\right)^2-4\right] +  1 \,.
\end{equation}

For the purpose of quantisation, we redefined  partially the phase
space variables \cite{Bergeron2015short,Bergeron2015long} of the isotropic geometry (scale factor $a=e^\Omega$) by introducing the canonical pair $(q,p) :=
(a^{3/2}, \;2 p_a/(3 \sqrt{a})$. This leads to a more convenient
form of Hamiltonian \eqref{con}:
\begin{align}\label{h4}
\mathcal{C}&=  \frac{3}{16 } p^2+\frac{3}{4 } q^{2/3}
-{\mathcal H}_q ,
\end{align}
  where
\begin{align}
\label{h4a}{\mathcal H}_{q}  &= \frac{1}{12 q^2} (p_+^2+p_-^2)
+\frac{3}{4}q^{2/3} V(\bsb)
\end{align}
is the anisotropy Hamiltonian.

\subsection{The anisotropy Hamiltonian $\mathcal{H}_{q}$}
We proved in \cite{qb9f}  the discreteness of the spectrum of the quantum Hamiltonian $\hat{\mathcal{H}}_{q}$ originated by the ``exact" Bianchi IX anisotropic potential $V(\bsb)$, despite the existence of three non-confining canyons of this potential. These canyons could suggest the existence of some continuum spectrum, but it is not the case. Moreover it was also shown on the classical level \cite{Uggla1, Uggla2} that those canyons do not contribute either to the chaotic behaviour of  the system. This validates all possible implementations of approximations of the potential removing the three non-confining canyons.  \\

Furthermore, we (with our co-authors) have shown in \cite{qmixV2} that the anisotropy potential $V(\beta)$  in \eqref{h4a} can be decomposed into two parts, each corresponding to a different Toda potential.
Therefore, approximating Mixmaster with Toda system  should preserve key properties of the model.
In a supplementary step, we showed in \cite{qmixV2} that after applying a Weyl-Heisenberg integral quantisation procedure to that classical potential (instead of a canonical quantisation), one of the parts of the potential become dominant whereas the other one becomes negligible. In other words, we showed that in the first order of approximation:
\begin{equation}
V^0(\bsb)\approx \frac{D^{16}}{3}\left(e^{4\sqrt{3}\beta_-+4\beta_+}+e^{-4\sqrt{3}\beta_-+4\beta_+}+e^{-8\beta_+}\right),
\end{equation}
where $D=e^{2/\sigma^2}$ is reminiscent  of the applied quantisation procedure, where $\sigma$ stands for weight of the applied Gaussian distribution coming from Weyl-Heisenberg quantisation. One may introduce  new variables $q_1,~q_2,~q_3$ such that:
\begin{equation} \label{qbeta}
q_1-q_2=4\sqrt{3}\beta_-+4\beta_+,\quad q_2-q_3=-4\sqrt{3}\beta_-+4\beta_+,
\end{equation}
that lead to
\begin{equation}
V^0(q_1,q_2,q_3)\approx \frac{D^{16}}{3}\left(e^{q_1-q_2}+e^{q_2-q_3}+e^{q_3-q_1}\right)\equiv \frac{D^{16}}{3} V_T(q_1,q_2,q_3),
\end{equation}
where  $V_T$ is 3-body Toda lattice potential $V_T(q_1,q_2,q_3)=e^{q_1-q_2}+e^{q_2-q_3}+e^{q_3-q_1}$.

Therefore, after suitable position, momenta and time reparameterisation (see \cite{qmixV2}), the Bianchi IX quantum anisotropy Hamiltonian \eqref{h4a} may be written in first order of approximation as
\begin{equation}
\hat{H}_q^0= \frac{\mathbf{p}^2}{q^2} + K q^{2/3} V_{T}(\bsb),
\end{equation}
where $K$ is a fixed constant, $\mathbf{p} = -i \hbar \nabla_\beta$ and $V_T(\vec{\beta})=e^{4\sqrt{3}\beta_-+4\beta_+}+e^{-4\sqrt{3}\beta_-+4\beta_+}+e^{-8\beta_+}$.

Therefore, we have
\begin{equation}
\hat{H}_q^0 = K q^{2/3} \left[ -\frac{1}{2} \left(\frac{\hbar}{ q^{4/3} \sqrt{K/2}} \right)^2 \nabla_\beta^2 + V_{T}(\vec{\beta}) \right].
\end{equation}

On the other hand,  a periodic 3-body Toda lattice  is actually a system with 2 degrees of freedom,  due to periodicity condition. In the paper \cite{Ford} it was presented how to reduce  the classical 3-Toda Hamiltonian \eqref{ham3per} to a Hamiltonian for a 2 dimensional system via a suitable change of position and corresponding momenta variables and also time reparameterisation   {similar to transformation \eqref{qbeta}}. Therefore, the 3-Toda classical Hamiltonian \eqref{ham3per} might be written as well at the quantum level (canonical quantisation) in a two dimensional form as follows
\begin{equation}
\hat{H}_{T} = -\frac{\hbar^2}{2} \nabla_\beta^2 + V_{T}(\bsb),
\end{equation}
where we denoted by  $\bsb$ the new  configurations variables \eqref{qbeta}.
Therefore, we obtain that the quantum Hamiltonian $\hat{H}_q(q, \hbar)$ formally fulfils
\begin{equation}
\hat{H}_q(q, \hbar) = K q^{2/3} \hat{H}_{T}(\hbar/\eta_q) \quad \text{with} \quad \eta_q = q^{4/3} \sqrt{K/2},
\end{equation}
Thus, the semi-classical EKB quantisation of Toda can be easily extended to Toda-Bianchi IX, by replacing $\hbar$ by $\hbar/\eta$ in \eqref{sys1} and then by renormalising the obtained energies by $K q^{2/3}$.

\section{Summary}
We proposed a novel parameterisation of the Toda lattice,   {the new varable description}  simplified significantly calculation action integrals coming from semi-classical EKB quantisation method. The new variables helped to ged rid of a large part of the initial numerical problem, replacing it by explicit formula (even if a final numerical step is unavoidable). This reduces significantly the process of finding energy levels of a 3-particle Toda system.

Regarding the energy levels of the ``exact" quantum anisotropy Hamiltonian of Mixmaster a semi-classical quantization like EKB seems interesting since this potential is ``not so far" from Toda potential, and it is known that the semi-classical approach is reliable for Toda  system\cite{Mutsuyama}.

On a side note, the explored  similarity between the anisotropy potential and the Toda potential was already used on the classical level, to introduce the so-called disturbed Toda lattices \cite{Bogo}.  Nevertheless, due to the integrability of the Toda latices this similarity seems to be too small  to be useful in the context of the classical dynamics (although see \cite{BSz}). However, quantum mechanics and classical mechanics are very different. For example, the Helium atom made of a nucleus and two electrons is a chaotic system at the classical level, but reliable quantum approximations of this system can be obtained without making reference of the chaotic classical behaviour. So it is not unreasonable to look at quantum Toda approximations of quantum Mixmaster. Furthermore, the Weyl-Heisenberg  quantization procedure that we used  in \cite{qmixV2}  amplifies the relative contribution of the positive Toda potential that is shown dominate over the negative one.  Thus, the latter can be neglected in the first approximation and the approximate evolution of the anisotropic variables for a fixed volume of the universe becomes integrable.  The negative Toda potential can be treated as a quantum perturbation to the integrable dynamics.

\section{Acknowledgements}
The project is cofinanced by the Polish National Agency for Academic Exchange and PHC POLONIUM 2019 (Grant No. 42657QJ).

\appendix

\section{Analysis of the equation $x^3-3x = a$}
\label{appendix}
This 3rd order polynomial  admits three solutions only if $|a| < 2$, therefore let us define $a$ as
$$
a = 2 \cos \phi\;,
$$
where we can assume  without loss  $0 < \phi < \pi$. \\
The three solutions of the title equation are now explicit in terms of $\phi$ and can be written as $x_1 < x_2 < x_3$:
\begin{equation}
\label{solspoly}
x_1 = -2 \cos \frac{\pi - \phi}{3}, \quad x_2 = -2 \cos \frac{\pi+ \phi}{3}, \quad x_3 = 2 \cos \frac{\phi}{3}.
\end{equation}
Furthermore:
$$
\text{for:} \quad 0 < \phi < \pi,  \quad-\cos \frac{\pi - \phi}{3}< -\cos \frac{\pi+ \phi}{3} <\cos \frac{\phi}{3}\quad \text{and}\quad x_1 < x_2 <  x_3,
$$
which comes from the properties of the cosine function.

\end{document}